# Ultrafast diffusive cross-sheet motion of lithium through antimonene with a 2+1 dimensional kinetics

Andrey A. Kistanov,[a, c] Devesh R. Kripalani,[a] Yongqing Cai,[*b] Sergey V. Dmitriev,[c,d] Kun Zhou,[*a] and Yong-Wei Zhang[*b]

Layered two–dimensional (2D) materials like graphene are highly appealing for lithium battery applications owing to their high surface-volume ratios. However, a critical issue that limits their practical applications is the confined motion of lithium atoms within their van der Waal's gaps, which is the leading cause for battery failure due to severe clustering and phase separation. Here we demonstrate that antimonene, an exfoliatable 2D material with a high structural stability, exhibits a highly mobile cross-sheet motion owing to its unique structural features. The advent of the vertically permeable channels opens a new pathway of lithium besides the normal motion along the basal plane, rendering a 2+1 dimensional kinetics. Specifically, our first-principles calculations combined with the discrete geometry analysis revealed that the energy barrier for a lithium atom to diffuse across the antimonene sheet is as low as 0.36 eV, which can be further reduced to 0.18 eV under a tensile strain of 4%. These ultralow diffusion barriers across the sheet can open a new dimension for controlling the motion of lithium atoms, leading to a new paradigm for high-performance lithium batteries or inorganic solid-state lithium-ion conductors.

## 1 Introduction

With the growing popularity of mobile gadgets and electric vehicles, there are increasing demands for high-power-density and fast-charging-discharging lithium-ion batteries (LIBs). One of the challenges in developing high-performance LIBs is to identify appropriate electrode materials.[1–3] Owing to their high surface-volume ratio, layered two-dimensional (2D) materials are ideal for high-performance LIBs electrodes, with the capability of relieving the strain associated with volume variations during lithium (Li) uptake-release process.[4, 5] For instance, due to their broad electrochemical window and excellent electrical properties, graphite and its structural derivatives are extensively used as the anode materials in the current generation of LIBs.[6] However, a critical issue inherent to these 2D materials-fabricated electrodes is that their quasi-flat geometry restricts the motion of Li atoms largely within their van der Waal's interlayer gaps, causing severe Li clustering and phase separation, which in turn lead to the failure of LIBs.[7, 8] Previous theoretical calculations revealed that the Li phase separation in a graphene anode could significantly reduce the capacity even down to zero.[9]

Although various thermodynamic and kinetic factors may affect the Li clustering and phase separation,[10] lack of the cross-sheet diffusion of Li atoms in layered LIBs materials is perhaps the utmost important reason. Indeed, the rate performance and packing density are improved through creating ordered and small pores in 2D sheets,[11] and aligned pore channels in layered transition metal oxide electrodes could deliver much faster charge transport kinetics with a more than threefold higher area capacity than that of conventional electrodes.[12] Ordered mesopores perpendicular to graphene layers were also found to enable efficient ion adsorption and transport, leading to an ultrahigh initial discharge capacity up to 3535 mAh/g.[13] It is noted that although the electrochemical and kinetic performances of these materials have been improved through structural engineering, the complex fabrication procedures prevent their practical applications. In addition, mesoscale pores in structurally tailored 2D sheets often cause highly irreversible side reactions with electrolyte solution, giving rise to a lower Coulombic efficiency (< 50%).[13, 14] Hence, finding a 2D material that intrinsically allows fast interlayer diffusion of Li atoms can potentially address this critical issue, thus facilitating the use of 2D materials for high-performance electrodes in LIBs.

Here we show that antimonene, a single layer of bulk antimony, could be such a system which potentially allows cross-sheet motion of Li without the need for structural engineering like creation of vacancies or mesoholes. Along with a high carrier mobility[15–17] and a high capacity,[18] antimonene also has a good stability at ambient conditions[19, 20] and under strain.[21, 22] Recent breakthroughs in the fabrication and exfoliation of antimonene[23–25] render it attractive and promising for LIB applications.[26] Using first-principles calculations, herein we reveal that Li atoms possess an ultralow diffusion barrier (0.36 eV) across antimonene, comparable to the motion in the horizontal basal plane. The underlying origin for the ultralow cross-sheet barrier is attributed to the much larger Wigner-Seitz radius of atomic antimony (1.58 Å) amongst elements (carbon: 0.86 Å, sulphur: 1.16 Å, and phosphorus: 1.23 Å) in other common 2D materials. Our work suggests that antimonene, the first 2D material with fast out-of-plane Li diffusion, could be unique with respect to the design of new-generation LIB electrode materials and solid ionic conductors.

## 2 Methods

### 2.1 First-principles electronic structure calculations.

The spin-polarized density functional theory (DFT)-based first-principles calculations were performed by using the Vienna ab initio simulation package (VASP).[27] Perdew-Burke-Ernzerhof (PBE) was selected to consider the electron-ion and the electron exchange-correlation interactions. The van der Waals-corrected functional with Becke88 optimization (optB88) was used for treating the dispersive interactions during the noncovalent chemical functionalization of antimonene with Li atoms. All the structures were fully relaxed until the atomic forces and total energy were smaller than 0.01 eV/Å and $10^{-5}$ eV, respectively. A plane wave cutoff of 400 eV and a 4×4×1 k-mesh grid were used for all the calculations. A vacuum of ~20 Å perpendicular to the atomic plane was used to avoid the interaction with spurious replica images. The relaxed lattice constants of antimonene were $a = b = 4.07$ Å, and the buckling height $h = 1.66$ Å, which are consistent with the results of the recent studies.[28-30] The diffusion barriers were calculated using the CI-

NEB. The AIMD simulations were performed at room temperature (300 K) using the Nose-Hoover method with a time step of 1.0 fs.

**2.2 Discrete geometry analysis.**

The discrete geometry was formulated using the method of triangulations over a finite mesh of atomic positions [31-33] (as illustrated in Fig. S1, see SI). Each triangulated element comprises three adjacent atoms (i.e., 1, 2, and 3), for which their spatial relation was given by the directed edges $e_1$, $e_2$, and $e_3$, where $e_1 + e_2 + e_3 = 0$. We defined $Z_p^I = e_p \cdot e_p$, ($p = 1, 2,$ and 3), which yields the square of the shortest distance between atoms; a discrete analogue of the infinitesimal length $ds^2$. The variation in orientation between the normal vectors $\hat{n}_q$ and $\hat{n}_r$ was projected onto their common edge $e_p$ following $Z_p^{II} = (\hat{n}_r - \hat{n}_q) \cdot e_p$. Here, $\hat{n}_q$ was taken to be the mean of the normal vectors of triangulated elements, which immediately surround atom p, enclosing it by a uniquely defined polygon, as outlined by the dashed lines in Fig. S1. Another relevant quantity, known as the dual edge, was given by $e^*_p = e_p \times \hat{v}$, where $\hat{v}$ is the normal vector of the triangulation described by atoms 1, 2, and 3. The area of the triangulated element in its pristine (reference) condition is denoted by $A_0$, while that under arbitrary deformation is indicated by $A_1$. The metric tensor g and curvature tensor k can then be expressed as follows:

$$g = -\frac{1}{8A_0^2} \sum_{(p,q,r)} \left(Z_p^I - Z_q^I - Z_r^I\right) \mathbf{e}_p^* \otimes \mathbf{e}_p^* \qquad (1)$$

$$k = -\frac{1}{8A_1^2} \sum_{(p,q,r)} \left(Z_p^{II} - Z_q^{II} - Z_r^{II}\right) \mathbf{e}_p^* \otimes \mathbf{e}_p^* \qquad (2)$$

The resulting tensor **g** and tensor **k** are in the form of 3×3 matrices, with the parentheses ($p, q, r$) representing the set of three contributions (1, 2, 3), (2, 3, 1), and (3, 1, 2) included in the summation. The eigenvalues $\{k_1, k_2, 0\}$ of the curvature tensor k provide us with the principal curvatures $k_1$ and $k_2$ for each triangulation such that the mean curvature $H = (k_1 + k_2)/2$ and Gaussian curvature $K = k_1 k_2$ can be evaluated. The four invariants Tr(**g**), Det(**g**), H, and K corresponding to each atomic position are averaged over their respective values at all triangulated elements that share the same vertex. The buckling height $\hat{t}$ of each atom in the lower (upper) sub-layer based on its out-of-plane distance from the z-centroid of its three nearest neighbours located in the upper (lower) sub-layer was also determined.

**2.3 Cluster expansion method for screening the configurational energy.**

The identification of stable ground state configurations for different Li concentrations was achieved through the cluster expansion technique. Based on the cluster expansion method, each lattice site can be represented by a spin-like variable $\sigma_i$, where $\sigma_i = +1$ if site $i$ is occupied by a Li atom and $\sigma_i = -1$ if site $i$ is a vacancy.

Accordingly, the energy of each configuration with spin $\sigma = (\sigma_1, \sigma_2, \sigma_3, ...)$ can be cast into a generalized Ising Hamiltonian:

$$E(\sigma) = J_0 + \sum_i J_i \sigma_i + \sum_{j<i} J_{ij} \sigma_i \sigma_j + \sum_{k<j<i} J_{ijk} \sigma_i \sigma_j \sigma_k + \cdots = $$
$$= J_0 + \sum_\alpha J_\alpha \varphi_\alpha \qquad (3)$$

where indices $i$, $j$, and $k$ range over all occupation sites, $\sigma_i$ denotes the pseudospin configuration variable for the respective lattice site $i$, $J_0$ is a constant shift and $J_\alpha$ is the ECI of cluster $\alpha$ which can be determined by fitting DFT-computed energies of selected configurations using a least-squares method. A cluster $\alpha$ consists of a particular combination of lattice sites $\alpha = (i, j, k, …)$, such as a single point, pairs, triples, quadruplets, etc. The predictability power of the cluster expansion model is measured by the CV score. The fitting process was performed using the Alloy Theoretic Automated Toolkit (ATAT).[34] More computational details can be found in Supporting Information.

# 3 Results and discussion

## 3.1 Facile cross-sheet motion of Li species through an antimonene sheet

The interaction of antimonene with the Li atom by considering several possible sites above antimonene is examined. As shown in Fig. 1a (upper left panel), two strongly anchoring sites are identified: the top of a Sb atom in the lower-lying plane (site A) and the centre of the neighbouring hexagonal hollow (site B) with the adsorption energies $E_a$ of –1.65 (site A) and –1.77 eV (site B). The distances $d$ calculated from the Li atom to the adsorption sites of antimonene are 1.50 (site A) and 1.35 Å (site B). The relatively strong binding of the Li atom with antimonene is accompanied with a strong upward shift of the Fermi level due to the charge transfer from the Li atom to the sheet (see Figs. S1 in SI).

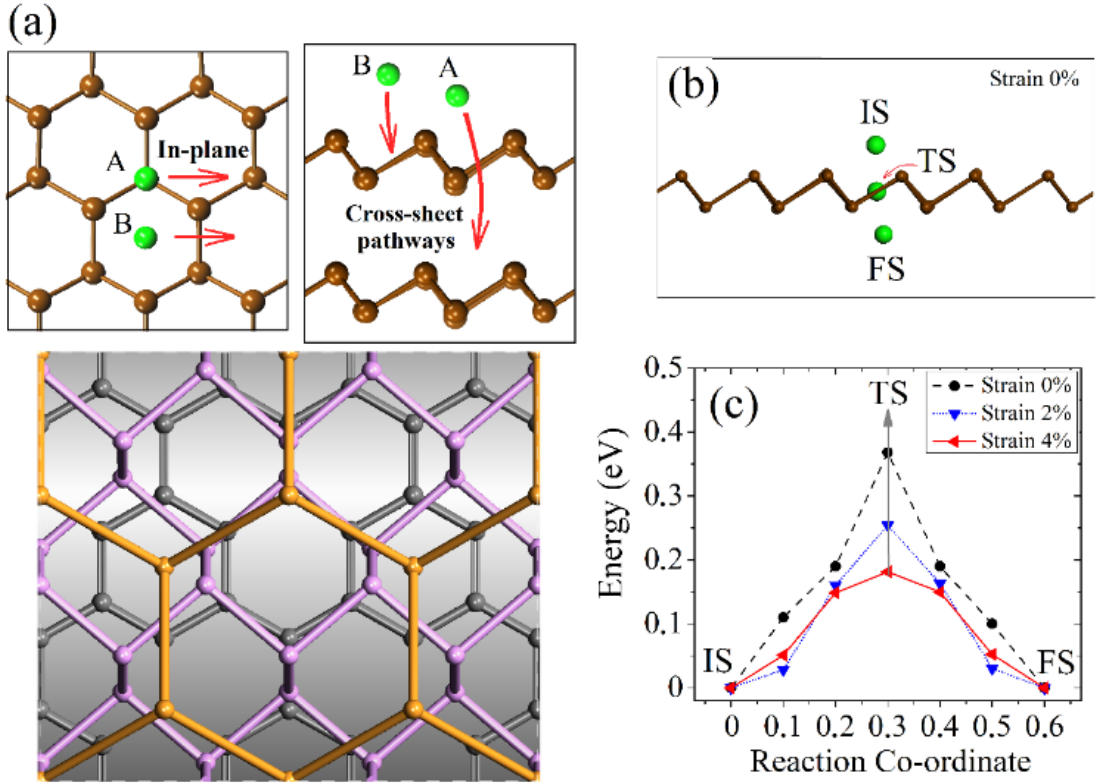

**Fig. 1.** (a) The upper panel: a schematic picture of the above-plane and cross-sheet pathways of Li diffusion in antimonene. The Li and Sb atoms are coloured in green and brown, respectively. The lower panel: a schematic picture of the graphene (grey), phosphorene (violet), and antimonene (brown) sheets. (b) The side views of the Li diffusion pathway through the antimonene surface both with and without strain. (c) The profiles of activation energy for cross-sheet diffusion of Li through the antimonene sheet under tensile strains.

We are particularly interested in the kinetic process of Li atoms on antimonene. Our calculations show that compared with other honeycomb elemental 2D materials, such as graphene and phosphorene, antimonene (Fig. 1a) has a similarly high Li mobility on antimonene. Herein, surprisingly, we demonstrate a novel cross-sheet pathway (Fig. 1a, upper-right panel) of Li in antimonene which is absent in other known 2D materials. Via climbing-image nudged elastic band (CI-NEB) calculation, we further analyse the diffusion energy barrier of the cross-sheet Li motion. Since the hopping event from site A to site B (Fig. 1a, upper right panel) is fast[26] and should not be a rate-limited process, we focus on the hopping process of the Li atom initially located at site B to the other side of the antimonene (Fig. 1b).

The schematic positions of the Li atom at the initial state (IS), transition state (TS) and final state (FS), and its corresponding energy profile are shown in Figs. 1b and c. Surprisingly, the barrier ($E_b$) of the cross-sheet motion through the freestanding antimonene is as low as 0.36 eV, implying a high transmitting probability of Li. This value is comparable to or even lower than that for the Li in-plane motion on the antimonene sheet, and other 2D materials like graphene (0.31 eV),[35] $MoS_2$ (0.52 eV),[36] and phosphorene (0.08 eV along the zigzag direction and 0.68 eV along the armchair direction).[37] In addition, the diffusion barrier for the cross-sheet Li motion through antimonene is also much lower than those of graphene[38] (10.2 eV), phosphorene[39] (1.19 eV) and $MoS_2$[40] (7.77 eV). Clearly, Li species adopt a new diffusive pathway through the cross-sheet transport in a highly mobile manner through the intrinsic spaces of the honeycomb lattice of antimonene, which has not been found in any other layered intercalated electrode materials so far. Interestingly, the low barrier of interlayer Li motion across antimonene suggests that antimonene cannot be exfoliated through the Li sonification as most of other 2D materials.

The underlying reason for this high diffusivity of Li along the cross-plane direction may be attributed to the large atomic radius of antimony. The Wigner-Seitz radius[41] of the antimony atom amounts to 1.58 Å which is double that of the carbon atom (0.86 Å) and much larger than those of the sulphur (1.16 Å) and phosphorus (1.23 Å) elements in other common 2D materials like graphene, $MoS_2$, and phosphorene. Therefore, the honeycomb lattice of antimonene contains much larger hexagons (Fig. 1a, lower panel), nearly four times those of graphene. This renders a much high permittivity of Li atoms across the antimonene layer. The geometrical origin of the low activation barrier is also reflected by the reduced $E_b$ of the cross-plane motion with biaxial tensile strain. As illustrated in Fig. 1c, $E_b$ is strongly affected by the tensile strain: $E_b$ decreases from 0.36 to 0.25 and to 0.18 eV with increasing the tensile strain from 0% to 2% and to 4%, respectively. Surprisingly, it turns out that the barrier ($E_b$ = 0.18 eV) for the cross-plane diffusion at 4% strain is comparable to that for the in-plane diffusion ($E_b$ = 0.11 eV,[29] no strain) of the freestanding antimonene sheet.

The hopping rate $v$ of a Li diffusing across the antimonene sheet is given by the Arrhenius formula[42]

$$v = v_s \exp(-E_b^*/kT) \quad (4)$$

where $k$ is the Boltzmann constant, $T$ is the temperature and $v_s$ is a prefactor. The $E_b^*$ is the diffusion barrier by taking account the zero-point energy correction (ZPE).[43,44] It should be noted that depending on the ambient working condition the zero-point energy corrections and tunneling corrections may play an important role in the estimation of the diffusion constant.[43] Recently, Guo et al.[44] have shown that the substantial quantum mechanical effect in the Li diffusion systems may be present at low temperatures.

The $v_s$ is defined by the Vineyard formula:

$$v_s = \frac{kT}{\hbar} \frac{\prod_{i=1}^{3N-3}(1-\exp(-\hbar\omega_i/kT))}{\prod_{i=1}^{3N-4}(1-\exp(-\hbar\omega_i'/kT))} \quad (5)$$

where $\omega_i$ ($\omega_i'$) is the atomic vibrational frequency of N atoms around the Li atom, which is at the initial or final (saddle) state. Due to the large size of the supercell, only the nearest neighboring atoms of the Li atom are considered in calculating the vibrational frequencies.

The $v$ for Li diffusion through antimonene surface at 300 K with and without ZPE correction are $1.26 \cdot 10^6$ Hz and $1.86 \cdot 10^6$ Hz, respectively, thus showing the importance of ZPE correction.

The temperature-dependent diffusion constant $D$ with ZPE correction for the cross-sheet hopping can be estimated as follows

$$D \approx v \cdot \frac{l^2}{2} \quad (6)$$

where $v$ is the hopping rate of a Li atom diffusing across the antimonene sheet and $l$ is the length of the hop. The $D$ with ZPE correction for Li diffusion through antimonene surface at 300 K is $\sim 0.13 \cdot 10^{-8}$ cm$^2$/s.

Table 1 compiles the hopping probability for Li diffusion in antimonene and other common 2D materials, such as graphene, MoS$_2$, and phosphorene. It is seen that antimonene possesses a high *in-plane* Li mobility, which is around $2 \cdot 10^4$ and $5 \cdot 10^3$ times faster than those of graphene and MoS$_2$, respectively.

Table 1. Comparison of the diffusion constant $D$ and average voltage of graphene, MoS$_2$, phosphorene, and antimonene.

| Material | Diffusion constant $D$ (cm$^2$/s) | | Average voltage (V) |
|---|---|---|---|
| | In–plane | Interlayer | |
| graphene | $3.2 \cdot 10^{-6}$ [45] | – | 0.20 [48] |
| MoS$_2$ | $6.3 \cdot 10^{-5}$ [46] | – | 0.67 [49] |
| phosphorene | $3.1 \cdot 10^{-2}$ [47] | – | 2.90 [37] |
| antimonene | $1.2 \cdot 10^{-2}$ [26] | $0.13 \cdot 10^{-8}$ (present work) | 3.0 (present work) |

**3.2 Discrete geometry analysis (DGA) of the Li-trapped antimonene** In this section, the deformation of the antimonene lattice associated with vertical intercalation of a Li atom is considered. The local atomistic structure of freestanding monolayer antimonene during cross-plane Li diffusion is analysed using discrete geometry within the context of 2D material' nets.[50, 51] As opposed to conventional continuum-based methods for parameterizing the strain field,[52] DGA goes beyond first-order elasticity to provide an exact description of shape under arbitrary deformation. This approach has facilitated the understanding of strain gauge fields and spin diffusion in graphene[53] as well as offered new insights into the role of strain and surface planarity in the chemical properties of various 2D systems.[31] The discrete geometry is characterized by four invariants of its metric tensor $g$ and curvature tensor $\mathbf{k}$, denoted by Tr($\mathbf{g}$), Det($\mathbf{g}$), mean curvature $H$ and Gaussian curvature $K$.[31] The reference values of these invariants are Tr($\mathbf{g}$) = 1, Det($\mathbf{g}$) = 1, $H$ = 0, and $K$ = 0, which correspond to the case of pristine antimonene in its strain-free planar configuration. The invariants Tr($\mathbf{g}$) and Det($\mathbf{g}$) take on values greater (less) than 1 for varying degrees of in-plane tension (compression), while $H$ and $K$ measure the out-of-plane deviation from planarity and character of the surface profile (i. e. ($K > 0$) for elliptical and ($K < 0$) for hyperbolic). Due to the buckled nature of antimonene, the normalized buckling height $\bar{t}$ provides another quantity of interest for our analysis. Note that, the discrete geometry is evaluated for the upper and lower sub-layers individually.

The discrete geometry of freestanding monolayer antimonene in its initial and saddle-point configurations during cross-plane Li diffusion is shown in Fig. 2, where changes in the local curvature ($H$ and $K$) are found to be negligible, and hence not presented. The initial configuration experiences moderate tensile strain in the lower sub-layer, while its buckling height decreases lattice smoothens out significantly with tripod-like symmetry. It is found that the presence of diffused Li within antimonene, as shown in the saddle-point configuration, induces even higher levels of tensile strain in both the upper and lower sub-layers. Here, the deformation is sub-layer symmetric, and indicative of the curvature-free interplay between in-plane strain and out-of-plane buckling during the diffusion process.

The introduction of strain, especially via heavy trapping of Li atoms, can have strong implications on the electronic properties of antimonene such as its work function and band structure. Notably, an indirect-to-direct band gap transition may be facilitated under small tensile strain (~4%) in the armchair direction.[23, 30, 53] This raises interesting opportunities for optimizing charge injection and transport across Li-intercalated antimonene contacts, paving the way forward for antimonene-based electrode materials in next-generation LIB applications.

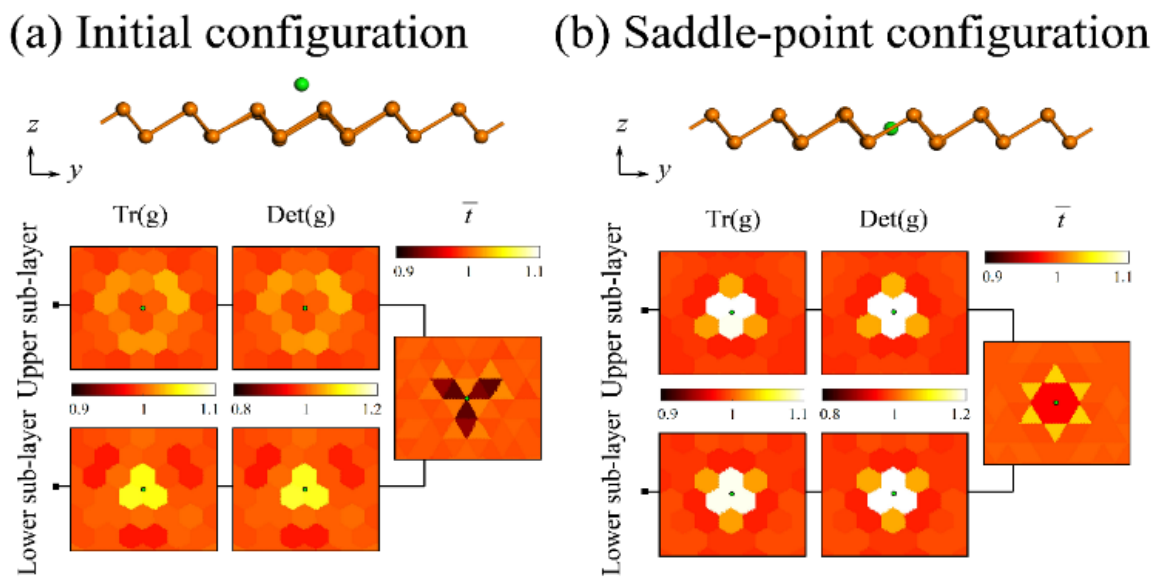

**Fig. 2.** The discrete geometry of freestanding monolayer antimonene during cross-plane Li diffusion for its (a) initial configuration, and (b) saddle-point configuration. The position of the Li atom relative to the deformation field is indicated by a green circle in the various sub-plots for Tr(g), Det(g), and $\bar{t}$.

### 3.3 Kinetics of the Li atomic diffusion through antimonene at room temperature

By using *ab-initio* molecular dynamics (AIMD) calculations, the kinetics of the Li atomic diffusion through antimonene is considered. The trajectories of the Li atoms on freestanding bilayer antimonene (Fig. 3) clearly show that several Li atoms randomly bounce up and down through the antimonene layer in a short period of time (~9 ps), which is also supported by the simulated snapshots in Figs. S3 and movie S1 (see SI).

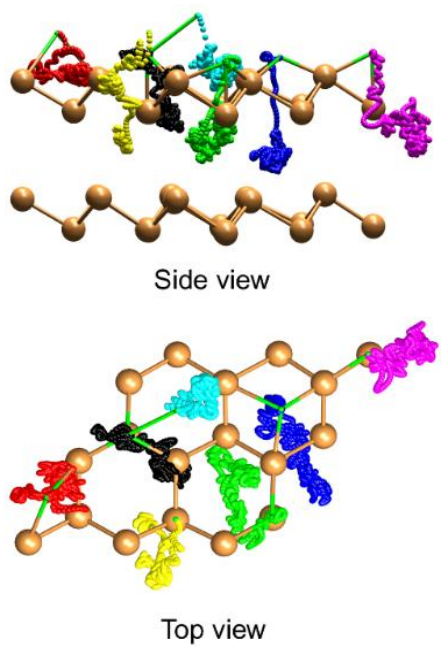

**Fig. 3**. The trajectories of the Li atomic motion through bilayer antimonene by AIMD simulations at 300 K. The Sb and Li atoms are coloured in brown and green, respectively. Diffusion pathways of Li atoms are represented by different colours.

The onset of Li atom diffusion through the top layer starts after 0.7 ps. Thereby, our findings confirm the ultrafast Li diffusion through the layered antimonene. Importantly, no clustering of Li atoms during the simulation is observed. In addition, AIMD study is performed to compare Li diffusion on freestanding (Figs. S4 and S5a–c and movie S2 in SI) and stretched monolayer antimonene (Figs. S5d–f and movie S3 in SI). It is seen that a small tensile strain (up to 4%) can substantially speed up (~3 times) the cross-sheet diffusion of the Li atoms through antimonene.

### 3.4 Configurational energetics of the antimonene lithiation

To search for the most energy-favourable Li configurations at an intermediate Li concentration, the cluster expansion formalism is adopted by evaluating the configurational dependence of the structural energies through the aid of DFT calculations. Previous works on the adsorption of Li adatoms on 2D substrates such as graphene, $MoS_2$, and phosphorene have shown that Li clustering and bulk-like coagulation becomes imminent as the concentration of Li increases.[37, 55] For antimonene, we also find that Li clustering occurs if the Li concentration is higher than 0.375, which is only slightly lower than that of graphene[48] (x < 0.5 in $Li_xC_6$) but higher than that of phosphorene[37] (0.25). Based on that, only moderate concentrations of Li, which corresponds to the chemical formula $Li_{3x}Sb_8$ ($0 \leq x \leq 1$), are considered. During the cluster expansion-based ground state search, we account for (i) the possibility of two-sided lithiation of antimonene and (ii) the availability of the two most favourable, hexagonal centre and buckling, sites for Li adsorption.

The effective cluster interactions (ECIs) are determined from a total energy fit to a training dataset containing 93 possible lithiated structures of different shapes, sizes, and concentrations. The predictive power of the resulting cluster expansion is verified from its low cross-validation (CV) score of 6.2 meV, which indicates negligible errors between the cluster expansion-predicted and DFT-calculated energies. Pristine ($x = 0$) and fully lithiated ($x = 1$) antimonene provide the configurational upper and lower limits in the search. The formation energies of all structures are predicted from the cluster expansion, revealing ground states of concentrations $x = \{0, 0.33, 0.44, 0.78, 0.83, 0.89, 1\}$, as shown in Fig. 4a. These ground states correspond to the chemical formulae $Sb_8$, $LiSb_8$, $Li_2Sb_{12}$, $Li_7Sb_{24}$, $Li_5Sb_{16}$, $LiSb_3$, and $Li_3Sb_8$, respectively. The unit cells of all ground states (intermediate and limit-defined) are represented in the insets of Fig. 4a. According to our calculations, the two-sided adsorption is found highly favourable at all intermediate ground state concentrations, as indicated by the presence of Li-occupied buckling sites on both the upper and lower surfaces of antimonene. Furthermore, the surface configuration of Li adatoms is observed to be comprised of a combination of both hexagonal centre and buckle sites for all the identified ground states. To show the structural stability of the Li-adsorbed antimonene structures obtained by the cluster expansion method, the stability of the structure with the maximum Li concentration ($L_3Sb_8$) is additionally proved by AIMD calculations (see movie S4 in SI).

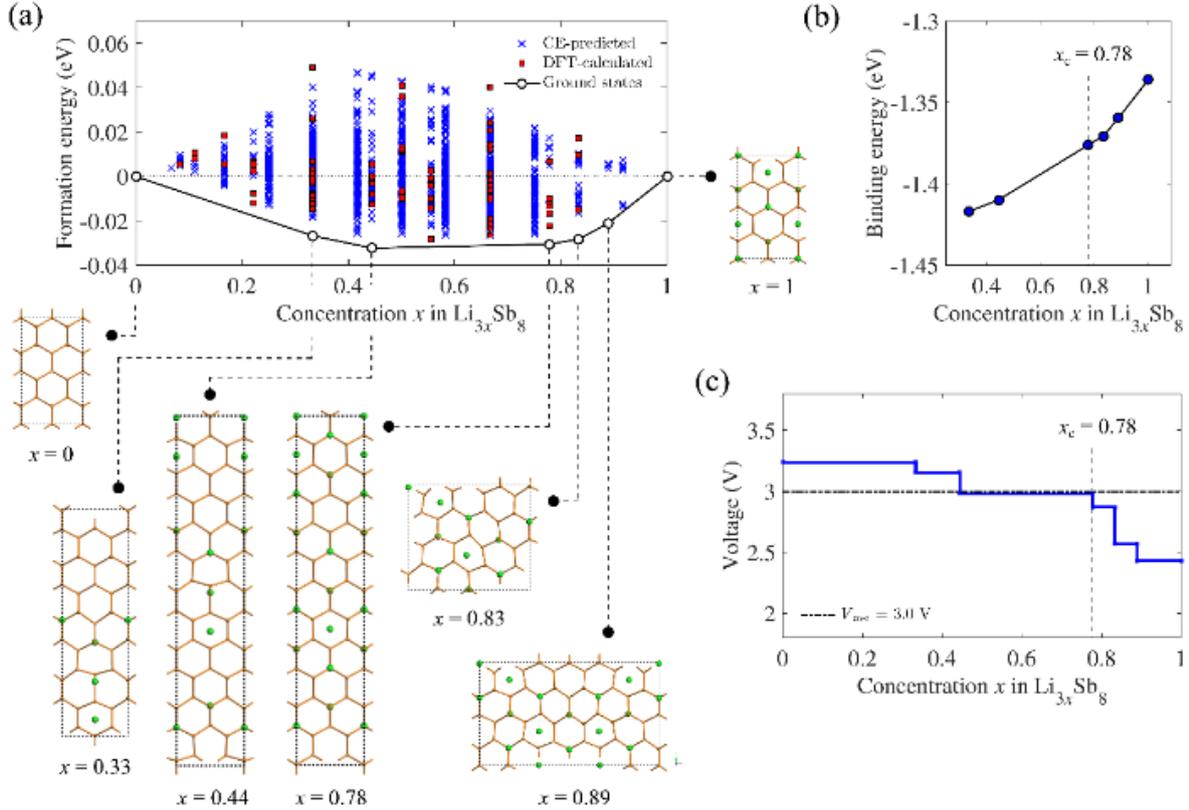

**Fig. 4.** (a) The formation energies of all the structures of $Li_{3x}Sb_8$ ($0 \leq x \leq 1$) predicted from the cluster expansion. The insets show the atomic structures of the ground states for $x = \{0, 0.33, 0.44, 0.78, 0.83, 0.89, 1\}$. (b) The binding energy as a function of the various ground states of $Li_{3x}Sb_8$ for $x = \{0, 0.33, 0.44, 0.78, 0.83, 0.89, 1\}$. (c) The open-circuit voltage of $Li_{3x}Sb_8$.

As shown in Fig. S6 (see SI), the buckling sites are primarily occupied at low concentrations up to a critical value of $x_c = 0.78$, beyond which the hexagonal centre sites become increasingly occupied. This general trend is consistent with our binding energy results corresponding to the various ground states, as shown in Fig. 4b. Here, a lower (higher) energy gradient is found on the left (right) side of $x_c$, which may be attributed to the stronger (weaker) binding contributions of the buckling (hexagonal centre) sites.

The binding energy (per atom) $E_{binding}$ reflects the relative stability of the various ground states and is given by

$$E_{binding} = \frac{1}{N_{Li}}[E_{Li+Sb} - (N_{Sb}E_{Sb} + N_{Li}E_{Li})] \quad (7)$$

where $N_{Li}$ and $N_{Sb}$ are the numbers of the Li and Sb atoms per unit cell, respectively. The energy term $E_{Li+Sb}$, refers to the total energy per unit cell of lithiated antimonene, while $E_{Sb}$ is the energy per atom of pristine antimonene, and $E_{Li}$ is the energy of an isolated Li atom. The $E_{binding}$ becomes less negative (i.e,. weaker binding strength) with increasing Li concentration.

The Another important property which has been widely used to assess the performance of materials with energy storage capabilities is the open-circuit voltage. In the context of our study, the mean voltage $V$ of $Li_{3x}Sb_8$ in the concentration range of $x_1 \leq x \leq x_2$ is given by[56]

$$V = -\frac{E_{Li_{3x_2}Sb_8} - (E_{Li_{3x_1}Sb_8} + (x_2-x_1)E_{Li-bcc})}{(x_2-x_1)e} \quad (8)$$

where $E_{Li_{3x_1}Sb_8}$ and $E_{Li_{3x_2}Sb_8}$ are the energies per formula unit of $Li_{3x_1}Sb_8$ and $Li_{3x_2}Sb_8$, respectively, while $E_{Li-bcc}$ is the energy per atom of bulk body-centered cubic (bcc) Li. The constant $e$ denotes the elementary charge quantity. As shown in Fig. 4c, the calculated open-circuit voltage of $Li_{3x}Sb_8$ decreases with increasing Li content in the system and spans between ~2.4 and 3.2 V. At low to moderate Li concentrations ($0 \leq x \leq x_c$), relatively gentle voltage drops are encountered, while sharper drops are observed at concentrations beyond the critical value $x_c = 0.78$. The average voltage $V_{ave}$ across the entire concentration range of interest is calculated to be 3.0 V, which is comparable to that of lithiated phosphorene (2.9 V) and even higher than those of graphene (~0.2 V), $MoS_2$ (0.67 V) (see Table 1), and $TiO_2$ (~1.5 V).[57]

## Conclusions

The diffusion and intercalation of Li in electrode materials are critical to the overall performance of the Li batteries. For traditional electrodes made of layered oxides or 2D materials, achieving a high-rate cross-sheet diffusivity of Li requires structural engineering, often by intentional introduction of atomic vacancies[58] or mesoholes.[11] This, however, may significantly complicate the design and leads to cyclability issues due to the coupling of electrolyte with the dangling defects. In this work, we demonstrate that antimonene is a unique 2D material that exhibits an ultrafast cross-sheet motion of Li species. Despite its 2D in nature, antimonene possesses a large void at the puckered hexagon centre, which offers a fast tunnelling pathway of Li along the out-of-plane direction, allowing for 2 + 1 dimensional kinetics of Li atoms in antimonene layers.

Both the cluster expansion and DFT-based calculations are performed to disclose the Li adsorption and diffusion in antimonene. The following conclusions can be drawn: (*i*) Li atoms on antimonene undergo ultrafast cross-sheet diffusion with an energy barrier of 0.36 eV, which is only slightly higher than that of the in-plane diffusion; (*ii*) moderate tensile strains can further reduce the barrier (0.18 eV at 4% tensile strain), which suggests that the deformation associated with the concave/convex regions of structurally rippled antimonene could favour the shuttling of Li species; (*iii*) DGA is performed to analyse the strain gauge fields associated with the Li insertion or extraction, which shows a curvature-free interplay between in-plane strain and out–of–plane buckling; (*iv*) AIMD results show that Li atoms are able to diffuse through the surface and back in monolayer and bilayer antimonene even at room temperature, suggesting a full freedom for the Li diffusion in antimonene; (*v*) the cluster expansion predictions suggest an impressive Li accumulation ability of antimonene (3 Li atoms per 8 Sb atoms) and a high charging voltage of 3.2 eV.

Thereby, our work suggests that antimonene, with its ultrahigh cross-sheet hopping rate of Li and high stability at ambient conditions, may be promising as an anode material for new-generation LIBs or ionic conductors.

## Conflicts of interest

There are no conflicts to declare.

## Acknowledgements

The authors acknowledge the financial support from the Agency for Science, Technology and Research (A*STAR), Singapore, and the use of computing resources at the National Supercomputing Centre, Singapore. This work was supported in part by a grant from the Science and Engineering Research Council (152–70–00017) and the state assignment of IMSP. A. A. Kistanov thanks the Russian Foundation for Basic Research, grant No. 18-32-20158 mol_a_ved. D. R. Kripalani acknowledges the support of the Economic Development Board, Singapore and Infineon Technologies Asia Pacific Pte Ltd through the Industrial Postgraduate Programme with Nanyang Technological University, Singapore.

## References

1  H. Li, Z. Wang, L. Chen and X. Huang, *Adv. Mater.,* 2009, **21**(45), 4593–4607.
2  M. Yang, Y. Zhong, J. Ren, X. Zhou, J. Wei and Z. Zhou, *Adv. Energy Mater.,* 2015, **5**(17), 1500550.
3  S. P. Ong, et al. *Energy Environ. Sci.*, 2011, **4**(9), 3680–3688.
4  L. Su, Y. Jing and Z. Zhou, *Nanoscale,* 2011, **3**(10), 3967–3983.
5  Y. Wang, W. D. Richards, S. P. Ong, L. J. Miara, J. C. Kim, Y. Mo and G. Ceder, *Nat. Mater.,* 2015, **14**(10), 1026–1031.
6  J. B.Goodenough and K. S. Park, *J. Am. Chem. Soc.,* 2013, **135**(4), 1167−1176.
7  S. J. Harris, A. Timmons, D. R. Baker and C. Monroe, *Chem. Phys. Lett.*, 2010, **485**, 265−274.
8  J. M. Tarascon and M. Armand, *Nature,* 2001, **414**(6861), 359−367.
9  Y. Liu, V. I. Artyukhov, M. Liu, A. R. Harutyunyan and B. I. Yakobson, *J. Phys. Chem. Lett.,* 2013, **4**(10), 1737−1742.
10  M. Liu, A. Kutana, Y. Liu and B. I. Yakobson, *J. Phys. Chem. Lett.*, 2014, **5**(7), 1225−1229.
11  J. Liu, S. Wang and Q. Sun, *Proc. Natl. Acad. Sci.,* 2017, **114**(4), 651−656.
12  J. S. Sander, R. M. Erb, L. Li, A. Gurijala and Y. M. Chiang, *Nat. Energy,* 2016, **1**(8), 16099.
13  Y. Fang, et al *J. Am. Chem. Soc.,* 2013, **135**(4), 1524−1530.
14  V. Etacheri, C. Wang, M. J. O'Connell, C. K. Chan and V. G. Pol, *J. Mater. Chem. A,* 2015, **3**(18), 9861–9868.
15  G. Pizzi, M. Gibertini, E. Dib, N. Marzari, G. Iannaccone and G. Fiori, *Nat. Commun.,* 2016, **7**, 12585
16  S. Zhang, et al. *Nano Lett.,* 2017, **17**(6), 3434–3440.
17  G. Wang, R. Pandey and S. P. Karna, *ACS Appl. Mater. Interfaces,* 2015, **7**(21), 11490−11496.
18  W. Tian, et al. *ACS Nano,* 2018, **12**(2), 1887–1893.
19  P. Ares, et al. *Adv. Mater.,* 2016, **28**(30), 6332–6336.
20  A. A. Kistanov, Y. Cai, D. R. Kripalani, K. Zhou, S. V. Dmitriev and Y. W. Zhang, *J. Mater. Chem. C,* 2018, **6**(15), 4308–4317.
21  H. Shu, Y. Li, X. Niu and J. Y. Guo, *J. Mater. Chem. C,* 2018, **6**(1), 83−90.
22  A. X. Zhang, J. T. Liu, S. D. Guo and H. C. Li, *Phys. Chem. Chem. Phys.,* 2017, **19**(22), 14520−14526.
23  S. Zhang, Z. Yan, Y. Li, Z. Chen and H. Zeng, *Angew. Chem. Int. Ed.*, 2015, **54**(10), 3112−3115.
24  J. Ji, et al. *Nat. Commun.,* 2016, **7**, 13352.

25  X. Wu, et al. *Adv. Mater.,* 2017, **29**, 1605407.
26  A. Sengupta and T. Frauenheim, *Mater. Today Energy,* 2017, **5**, 347−354.
27  G. Kresse and J. Furthmüller, *Phys. Rev. B Condens. Matter.,* 1996, **54**(16), 11169–11186.
28  S. Zhang, M. Xie, Z. Yan, E. Kan, W. Liu and H. Zeng, *Angew. Chem.*, 2016, **128**, 1698–1701.
29  S. Zhang, S. Guo, Z. Chen, Y. Wang, H. Gao, J. Go´mez-Herrero, P. Ares, F. Zamora, Z. Zhu and H. Zeng, *Chem. Soc. Rev*., 2018, **47**, 982–1021.
30  D. R. Kripalani, A. A. Kistanov, Y. Cai, M. Xue and K. Zhou, *Phys. Rev.B*, 2018, **98**, 085410.
31  M. Mehboudi, K. Utt, H. Terrones, E. O. Harriss, A. A. P. Sanjuan and S. Barraza−Lopez, *Proc. Natl. Acad.,* 2015, **112**(19), 5888−5892.
32  C. Weischedel, A. Tuganov, T. Hermansson, J. Linn and M. Wardetzky, The 2nd Joint Conference on Multibody System Dynamics Stuttgart, Germany, **2012**.
33  A. Bobenko, P. Schroder, J. Sullivan and G. Ziegler, *Eds discrete differential geometry*. 1st ed Oberwolfach Seminars Birkhauser: Basel, Switzerland, **2008**.
34  A. Van de Walle and G. Ceder, *J. Phase Equilib.,* 2002, **23**, 348.
35  X. Fan, W. T. Zheng and J. L. Kuo, *ACS Appl. Mater. Interfaces,* 2012, **4**, 2432−2438.
36  B. Xu, L. Wang, H. J. Chen, J. Zhao, G. Liu and M. S. Wu, *Comp. Mater. Sci.,* 2014, **93**, 86–90.
37  W. Li, Y. Yang, G. Zhang and Y. W. Zhang, *Nano Lett.,* 2015, **15**(3), 1691–1697.
38  F. Yao, F. Günes, H. Q. Ta, S. M. Lee, S. J. Chae, K. Y. Sheem, C. S. Cojocaru, S. S. Xie and Y. H. Lee, *J. Am. Chem. Soc*., 2012, **134**, 8646−8654.
39  S. Zhao, W. Kang and J. Xue, *J. Mater. Chem. A*, 2014, **2**, 19046–19052.
40  M. Li, J. Luo and W. Wang, Proceedings of the 17th IEEE international conference on nanotechnology, Pittsburgh, USA, 2017.
41  L. A.Girifalco, Statistical mechanics of solids. Oxford: Oxford University Press. 2003, **125**.
42  S. A. Arrhenius, *Z. Phys. Chem*., 1889, **4**, 96–116.
43  S. Karmakar, C. Chowdhury and A. Datta, *J. Phys. Chem. C*, 2016, **120**, 14522−14530.
44  S. Guo, X, Hu, W. Zhou, X. Liu, Y. Gao, S. Zhang, K. Zhang, Z. Zhu and H. Zeng, *DOI: 10.1021/acs.jpcc.8b08824*.
45  G. Wang, X. Shen, J. Yao and J. Park, *Carbon,* 2009, **47**(8), 2049–2053.
46  R. Dominko, et al. *Adv. Mater.,* 2002, **14**(21), 1531−1534.
47  C. Zhang, M. Yu, G. Anderson, R. R. Dharmasena and G. Sumanasekera, *Nanotechnology,* 2017, **28**(7), 075401.
48  K. Persson, Y. Hinuma, Y. S. Meng, A. Van der Ven and G. Ceder, *Phys. Rev. B,* 2010, **82**(12), 125416.
49  J. Hao, et al. *Sci. Rep*., 2018, **8**, 2079.
50  J. Lee, Riemannian manifolds: An introduction to curvature. *Springer−Verlag* New York, 1997, *Vol. **176**,* pp 437.
51  A. I. Bobenko and Y. B. Suris, *Amer. Math. Soc.* 2008, **98**, 404.
52  F. de Juan, A. Cortijo and M. A. H. Vozmediano, *Phys. Rev. B,* 2007, **76**(16), 165409.
53  M. Pumera and Z. Sofer, *Adv. Mater.*, 2017, **29**, 1605299.
54  J. V. Sloan, A. A. P. Sanjuan, Z. Wang, C. Horvath and S. Barraza−Lopez, *Phys. Rev. B,* 2013, **87**(15), 155436.
55  X. Fan, W. T. Zheng, J. L. Kuo and D. J. Singh, *ACS Appl. Mater. Interfaces,* 2013, **5**(16), 7793–7797.
56  D. Er, J. Li, M. Naguib, Y. Gogotsi and V. B. Shenoy, *ACS Appl. Mater. Interfaces*, 2014, **6**, 11173–11179.
57  Z. Yang, et al. *J. Power Sources,* 2009, **192**(2), 588–598.
58  A. Van der Ven, J. Bhattacharya, A. A. Belak, *Acc. of Chem. Res.,* 2012, **46**(5), 1216.